# Solar Ions In The Heliosheath: A Possible New Source Of Heavy Neutral Atoms

S. Grzedzielski[1], M. Wachowicz[1], M. Bzowski[1] and V. Izmodenov[2]

(1) Space Research Centre, Polish Academy of Sciences, Bartycka 18A, 00-716, Warsaw Poland
(2) Lomonosov Moscow State University, Department of Mechanics and Mathematics & Space Research Institute (IKI) Russian Academy of Sciences, Russia

**Abstract.** We show that multiply ionized coronal C, N, O, Mg, Si, S ions carried by the solar wind and neutralized by consecutive electron captures from neutral interstellar atoms constitute an important new source of neutral atoms in the inner heliosheath, with energies up to ~ 1 keV/n. In the model we developed, the heavy ions are treated as test particles carried by hydrodynamic plasma flow (with a Monte-Carlo description of interstellar neutrals) and undergoing all relevant atomic processes determining the evolution of all charge-states of considered species (radiative and dielectronic recombination, charge exchange, photo-, and electron impact ionization). The total strength of the source is from $\sim 10^6$ g/s for S to $\sim 10^8$ g/s for O, deposited as neutrals below the heliopause. These atoms should provide, as they drift to supersonic wind region, important sources of PUIs and eventually ACRs, especially for species that are excluded from entering the heliosphere because of their ionization in the LISM. The expected corresponding ENA fluxes at 1 AU are in the range $10^{-4} - 10^0$ at./(cm$^2$ s sr), depending on the species and direction (Table2).

## INTRODUCTION

Opinion prevails that heavy neutral atoms in the heliosphere come from either (i) interstellar space or (ii) solar system objects (from dust to planets). Ionization of these atoms provides the source for the observed pickup and anomalous cosmic ray ion populations (PUI & ACR). However, neutralization of multiply charged heavy solar wind ions by consecutive electron captures from neutral interstellar hydrogen can also be a source of heavy neutral atoms, and therefore of PUI & ACR. To evaluate this additional source we follow the evolution of heavy ion content of solar wind parcels as they move along their flow lines from inside Earth orbit to distant solar wind, to inner heliosheath and finally to heliospheric tail, and we calculate the net effect of all atomic processes that could have affected on the way the charge states of solar C, N, O, Mg, Si, and S ions. We show in this paper that the strength of this additional source is very significant and for some low FIP (first ionization potential) species may be dominant in the heliosphere.

# BASIC FEATURES OF THE MODEL

We consider multiply ionized heavy ions of ~1 keV/nucleon in the solar wind to be test particles carried by the general flow of interplanetary plasma. They undergo binary interactions (collisions) with solar wind electrons, protons, with solar ionizing photons, as well as with neutral interstellar hydrogen atoms (in some calculations neutral helium atoms were also included). As interactions we take: radiative and dielectronic recombinations, impact ionizations, photoionizations and charge exchanges. Among all these the most important, especially for multiply charged ions, is electron capture from neutral hydrogen. Upon a single interaction the charge carried by the heavy ion may change, i.e. increase or decrease by one, depending on the process. Therefore, the time evolution along the flow line of the number $N_{(Z,A)}^{+i}$ of ions (originating from atoms Z,A), carrying charge +i (i = 0,..., Z) and contained within a solar wind parcel of unit mass, can be described by the equation

$$d\, N_{(Z,A)}^{+i}/dt = \Sigma(\text{recombinations}) + \Sigma\,(\text{ionizations}) + \Sigma\,(\text{charge exchanges}), \quad (1)$$

where the summations are over rates per unit mass per s for all relevant processes. To calculate the RHS of eq.(1) one should know, besides the cross sections, the background solar plasma and neutral interstellar gas flows, i.e. the spatial distributions of solar wind electron ($n_e$) and proton ($n_p$) densities, the density distribution of interstellar neutral hydrogen ($n_H$) as well as the corresponding bulk flow velocities of the solar wind ($\mathbf{v}_{sw}$) and interstellar hydrogen ($\mathbf{v}_H$) and the effective relative velocities of particles at collisions ($v_{rel}$) resulting from (local) particle velocity distribution functions.

The background flows of solar plasma and neutral hydrogen atoms in supersonic solar wind, inner heliosheath and distant heliospheric tail were calculated by [1] basing on the time-independent, single-fluid, non-magnetic, gasdynamical model for plasma coupled by mass, momentum and energy exchange with neutral interstellar hydrogen atoms, as originally developed by [2]. In this self-consistent treatment the neutral H distribution was calculated kinetically (Monte-Carlo). The Sun, as source of solar wind and ionizing photons, is assumed to be spherically symmetric, with the wind speed of 450 km/s, Mach number 10 and $n_p$ = 7 cm$^{-3}$ at Earth orbit. At infinity, a uniform interstellar flow of 25 km/s, with $n_{H,LISM}$ = 0.2 cm$^{-3}$, $n_{p,LISM}$ = 0.07 cm$^{-3}$ and temperature 6000 K was assumed.

In view of axial symmetry, all variables depend on the radial distance r from the Sun and angle θ from the apex direction. The evolution of spatial density of all ionization states for the six selected species was calculated by numerical integration of eq. (1) along 180 flow lines, corresponding to initial values of $\theta$ = 1,2,...180 degrees (at Earth orbit). Using $|\mathbf{v}_{sw}|dt = ds$, the time integration of eq.(1) can be transformed into a space integration along the curvilinear coordinate s running along each of the flow lines. In this way a complete spatial distribution of all $N_{(Z,A)}^{+i}$ for every considered species can be obtained.

As long as the solar wind parcel moves supersonically between the Sun and the termination shock, the heavy ions can be thought to adiabatically cool similarly to the background plasma and therefore stay in approximate thermal equilibrium with the local particle environment. In this case $v_{rel}$ for all interactions with electrons (radiative and dielectronic recombination, ionizing impacts) and for charge exchange reactions with protons was calculated assuming particle velocity distributions to be maxwellians corresponding to local (single-fluid) temperature. For heavy ion – neutral atom interactions $v_{rel}$ = solar wind bulk speed = 500 km/s was taken.

Upon crossing the termination shock the proton-electron plasma on a single-fluid model heats up to about ~$10^6$ kelvin. However, as the heliosheath cooling times for ~1 keV/n heavy ions by Coulomb scattering on protons and by collisions with H (~$10^{11}$ s for $O^{+8}$, and longer for lesser charges and higher masses) are much longer than heliosheath flow times ($10^8 - 10^9$ s), we assume the post-shock heavy ions to isotropize velocities but retain their energies in the heliosheath. Therefore, while in this region we use the 'hot' maxwellian rates to describe the various plasma-plasma collisions, the $v_{rel}$ values for heavy ion – neutral atom collisions are still ~500 km/s. (We add here that calculations corresponding to full or partial post-shock thermalization of heavy ions are now also under development). A significant effort was made to collect adequate cross sections, in particular for electron capture by multiply charged ions [3]. The initial values of $N_{(Z,A)}^{+i}$ were taken at Earth orbit basing on the results of MTOF/Celias on SOHO, SWICS on Ulysses and SWIMS on ACE[4],[5],[6]. It was tested that the results are not sensitive to details of initial conditions. In effective calculations initial $N_{(Z,A)}^{+i}$ values based on SWICS data were used (Table 1).

**TABLE 1**. SWICS Initial Fractional Ion Abundances

| C | | N | | O | | Mg | | Si | | S | |
|---|---|---|---|---|---|---|---|---|---|---|---|
| +4 | .6 | +4 | .4 | | | +6 | .02 | +7 | .06 | +8 | .21 |
| +5 | .3 | +5 | .4 | +6 | .67 | +7 | .13 | +8 | .17 | +9 | .29 |
| +6 | .1 | +6 | .2 | +7 | .30 | +8 | .15 | +9 | .43 | +10 | .36 |
| | | | | +8 | .03 | +9 | .33 | +10 | .18 | +11 | .14 |
| | | | | | | +10 | .37 | +11 | .10 | | |
| | | | | | | | | +12 | .06 | | |

Also addition of constant neutral helium background with $n_{He}$ = 0.015 cm$^{-3}$ [7] did not change the results by more than 10%. The integrations were carried deep into the heliospheric tail (in some cases to about $4 \times 10^4$ AU).

## DISCUSSION OF RESULTS

Fig. 1 (left panel) shows the geometry of flow lines in the upwind and near tail heliosphere and, as an example, (right panel) the time evolution of relative abundances for all ionization states of oxygen (i.e. $N_{(8,16)}^{+i}$ divided by the total number of O-ions per unit wind mass) along a flow line that starts 30$^o$ off apex direction. Vertical red lines indicate times (in s) of termination shock and cross wind direction ($\theta = 90^o$) passage. Decreasing size of black dots corresponds to decreasing ion charge (largest – $O^{+8}$, smallest – $O^{+2}$). The green line describes the evolution of $O^{+1}$ abundance, in particular its rise as plasma moves away from the upwind heliosphere.

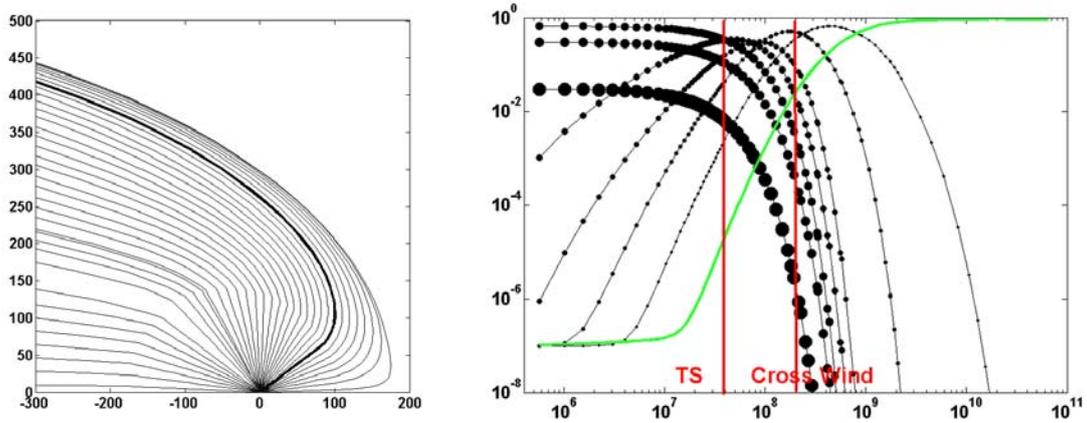

**FIGURE 1.** Left panel: geometry of flow lines in the upwind and near tail heliosphere (scale in AU). Only 30 out of 180 flow lines are shown. Right panel: time evolution (time in s) of relative abundance for all ionization states of oxygen for a flow line (solid on the left panel) starting at Sun 30° off apex direction. The green curve gives the relative abundance of $O^{+1}$ ions (see text for more explanations).

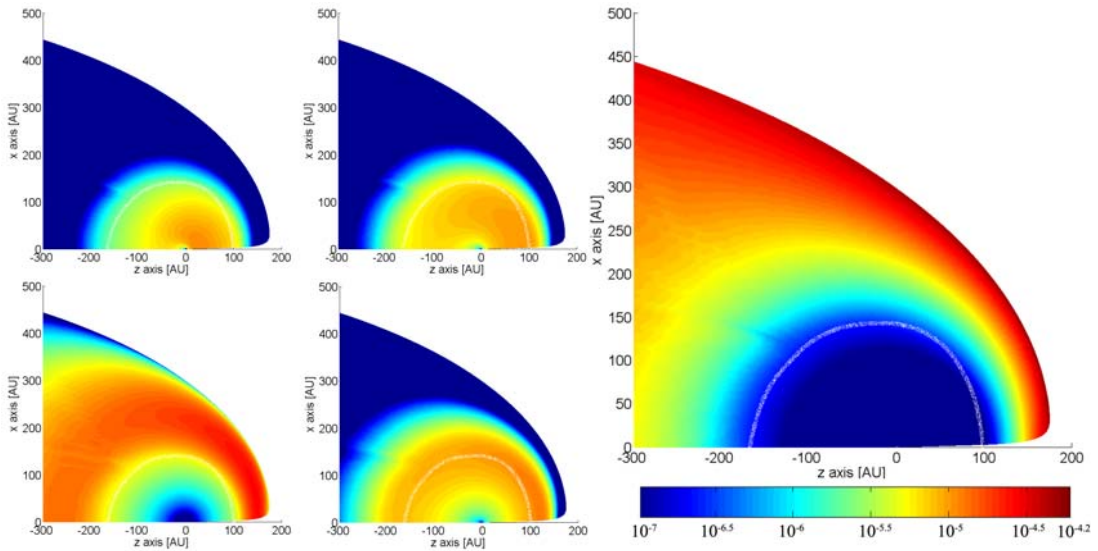

**FIGURE 2.** Left panel: (clockwise from upper left) $C^{+6}$, $C^{+5}$, $C^{+4}$, $C^{+3}$ densities in the heliosphere. Right panel: $C^{+1}$ density. Color coding gives density in cm$^{-3}$. White band denotes the termination shock.

As typical specimen of our results we present in Fig. 2 heliospheric maps of density distributions (ions/cm$^3$) of carbon ions in various ionization states. The left panel represents (clockwise from upper left) $C^{+6}$, $C^{+5}$, $C^{+4}$, $C^{+3}$. The right panel shows $C^{+1}$ density on an enlarged scale, with common color coding for ion density in cm$^{-3}$. Note a high density ridge appears for $C^{+3}$ beyond the termination shock. Finally, for $C^{+1}$ one obtains a strong density enhancement towards the heliopause.

Similar maps were obtained for all considered species. Common feature of all these ion distribution maps is that close to the heliopause a layer of relatively high density of singly ionized ions develops in the heliosheath. The typical density contrast between the maximum in that layer (which in most cases, though not all, is lining the heliopause) and the region adjacent to the termination shock is of the order of ~50, ~$10^3$, ~$10^4$, ~$10^3$, ~$10^4$, ~$3 \times 10^3$ for $C^{+1}$, $N^{+1}$, $O^{+1}$, $Mg^{+1}$, $Si^{+1}$, $S^{+1}$. Existence of this relatively high density layer for singly ionized species is a direct consequence of the fact that heliosheath flow time scales are longest for flow lines passing closest to the heliopause: this provides for ions more chance for electron capture from neutral gas. Differences between the maxima of various species reflect differences in the individual cross sections.

Singly ionized ions produce neutral atoms by charge exchange with interstellar H (and He). Integration of this production over the entire upwind heliosheath (to cross wind, $\theta = 90^o$) gives total neutral gas production rates (g/s) per species *inside* the upwind section of the cavity carved by the heliopause. The obtained rates (Table2) are significant by heliospheric standards, especially for low FIP species [8], [9]. These neutral atoms will easily drift over the entire heliosphere, in particular into the supersonic solar wind, and produce populations of PUI and ACR ions much like in the case of neutral atoms flowing directly from interstellar space.

**TABLE 2**. Neutral Atom Production in the Heliosheath and Expected ENA Fluxes

| Species | Total Neutral Gas Production To Cross Wind (g/s) | Expected Average ENA Fluxes in Distant Wind After Heliosheath Losses (cm$^2$ s sr)$^{-1}$ | Expected LISM Neutral Flux at Heliopause Nose (cm$^2$ sr)$^{-1}$ | Expected ENA Fluxes at 1 AU After Losses (cm$^2$ s sr)$^{-1}$ $\Theta = 0^o$ | $\Theta = 90^o$ |
|---|---|---|---|---|---|
| C  | $6.4 \times 10^7$ | 2.6   | 0.08      | 1.1    | 1.6 |
| N  | $1.1 \times 10^7$ | 0.4   | 18        | 0.2    | 0.3 |
| O  | $9.3 \times 10^7$ | 2.1   | 174       | 0.7    | 1.6 |
| Mg | $1.3 \times 10^7$ | 0.1   | 0.006     | 0.05   | 0.07 |
| Si | $1.4 \times 10^7$ | 0.005 | <<0.004   | 0.0002 | <$10^{-4}$ |
| S  | $1.1 \times 10^6$ | 0.01  | 0.01      | -      | - |

We suggest, therefore, that the described process constitutes a new important source for the PUIs and ACRs. Three important points should be stressed in this respect:
(1) in contrast to interstellar supply, the present mechanism deposits neutral atoms right inside the heliopause; thus are avoided difficulties associated with species with low FIP (C, Mg, Si, S) that should be fully ionized in the local interstellar medium and are therefore prevented from penetrating below the heliopause;
(2) the heliosheath production rates in Table 2 are significant by heliospheric standards when compared with possible external [10] (cf. Table 2) and "internal" sources for PUIs and ACRs; this should facilitate balancing the budget of production and losses for the energized ions;
(3) as long as the heavy ions in the heliosheath preserve, as we argued above, their original energies of ~1keV/n, the resulting neutral atoms will be of the ENA type with about the same energy, capable of penetrating deep into the supersonic solar wind (< 1 AU); the expected ENA fluxes at 1 AU from the upwind heliosphere are then of the order of $10^{-4} - 10^0$ at./(cm$^2$ s sr), depending on the species and direction (Table 2).

We conclude, therefore, that the new source of heavy atoms we predict to operate within the confines of the heliopause can be expected to play an important role in the physics of heliospheric PUIs, ACRs and ENAs.

## ACKNOWLEDGMENTS

This research was supported by Polish MSRiT Grant 1P03D 009 27.
V. I. acknowledges support of RFBR grants 04-02-16559, 05-02-22000-CNRS (PICS), "Dynastia" Foundation and "Foundation in Support of Russian Science".